# Superconductors:

# The Standing Wave Model and Superfluid Density

Refael Gatt

*Quantum Designed Materials Ltd.*

Abstract

The standing wave model describes the well-known phenomenon of superconductivity in a new way [1]. Starting from a new definition of superconductivity, a microscopic London relation is derived from first principles. The relation between the macroscopic electric current and the field is derived from this microscopic relation. The coherence length as function of the superfluid density is derived from the microscopic theory. The critical temperature as function of the superfluid density is thus obtained. These results are compared to the experimental work of Bozovic et. al. [2].

**Introduction**

Since the discovery of high temperature superconductors [3], several key experiments were performed, including μSR [4], ARPES [4,5], STM [6], inelastic neutron diffraction [7] and other experiments [8,9,10] in an effort to understand the mechanism of superconductivity in these materials and whether it can be described by the theory of BCS [11] or by some new theory [12, 13,14].

With 30 years of experimental and theoretical work, we are still lacking an understanding of the mechanism of high temperature superconductivity. In this paper, we show that the new results of reference [2] strongly support the standing wave model, a different way of understanding the phenomenon of superconductivity in general, and the interactions that are responsible for it.

The work of Bozovic et. al., [2] describe the results of 2000 MBE-grown $La_{2-x}Sr_xCuO_4$ samples measured by inductive coupling. Fig. 2 in their paper shows a unique dependence of Tc on the superfluid density with varying doping levels in the over-doped region of the phase diagram. As stated in the paper, no existing theory can explain the results. This short article describes calculations that reproduce the observed dependence of the critical temperature on the pair density for $La_{2-x}Sr_xCuO_4$. The calculations are based on the standing wave model [1].

The results of Bozovic et al. [2] were obtained by magnetic measurements on $La_{2-x}Sr_xCuO_4$ with varying doping levels. From the magnetic measurements, Bozovic et al. derived the penetration depth λ as function of temperature. When extrapolated to T=0K, the parameter $\rho_{s0}$ (the superfluid density at T=0K) was extracted from the value of λ at T=0K through the London relation [16]. Bozovic et. al. claim that no existing theory can explain their result.

### Superfluid Density in the Standing Wave Model

The Standing Wave Model (SWM) begins with a single assumption:

$$\nabla_k \varepsilon = 0 \big|_{\varepsilon=\varepsilon_0} \quad \text{in the superconducting state.} \tag{1}$$

This is a single electron relation, so it corresponds to the London description of super-electrons. Pairing correlation emerges as a many-body result of that single super-electron property [1]. The assumption (1) does not imply any long-range coherence. Such coherence is derived from this single assumption.

The electromagnetic properties of superconductors emerge as a direct consequence of that assumption. This is because a standing wave state produces a microscopic London relation between the single electron probability current and the vector potential [1].

The macroscopic electrical current is then derived as a collection of these microscopic currents, producing a Pippard-like relation between the current and the field. This relation dictates a certain form for the kernel connecting the current and the field. This specific form of the kernel allowed us to calculate the coherence length as function of the superfluid density from first principles and therefore the dependence of Tc on the superfluid density. This relation shows astonishing fit to the results of Bozovic *et. al.* on LSCO thin films.

Zero resistance, the Meissner effect, flux quantization and all other electrodynamic properties of superconductors are derived from this microscopic London relation and the macroscopic Pippard-like relation derived from it.

The standing wave model dictates the single assumption (1).

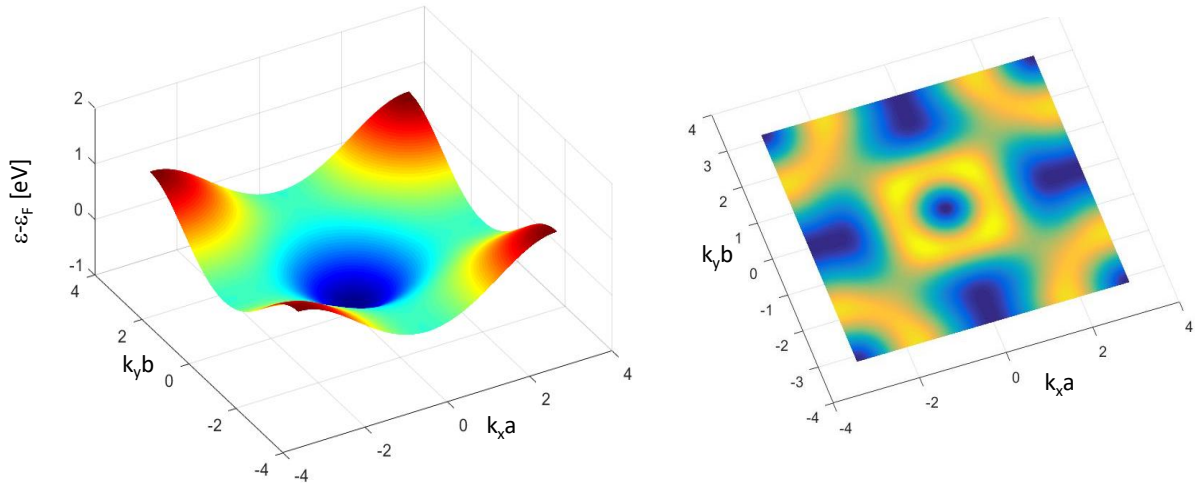

**Fig. 1 Left: The dispersion relations $\varepsilon(k)$ at the proximity to the Fermi level for optimally doped LSCO as measured by ARPES [15]. Right: The square of the gradient of $\varepsilon(k)$ at the same neighborhood of $\varepsilon(k)$ (color map).**

Therefore, only states with very shallow dispersion will participate in the pairing. (The pairing in the standing wave model is derived from equation (1) and is not imposed. For our purpose, we can consider even single super electron states as defined by London [16]). If that assumption is right, we should be able to estimate both the critical temperature (Tc) and the super-fluid density at T=0 ($\rho_{s0}$) from the dispersion relations at the proximity of the Fermi surface as they appear in Fig. 1 [17]. Fig. 1 shows the results of Angular Resolved Photoemission (ARPES) on $La_{2-x}Sr_xCuO_4$ at optimal doping [15]. The left panel shows a tight binding approximation for the measured $\varepsilon(\mathbf{k})$ dispersion relations at the proximity to the Fermi level. The right panel of Fig. 1 shows the map of $|\nabla_k \varepsilon|^2$ as function of $\mathbf{k}$. From this map, it is possible to compute within the standing wave model, the coherence length and the pair density, assuming a rigid band model. The coherence length $\xi_0$ was defined by Pippard [18] as the length scale of the decay of the kernel K(**r**-**r'**) connecting the electric current **J**(**r**) with the vector potential **A**(**r'**) in a pure material. In the standing wave model [1], the London relation [16] appears as a *microscopic* relation between the probability current of a single super-electron and the vector potential:

$$J(r,t) = Re\left[\psi(r,t)^* \left(\frac{\hbar}{im}\nabla - \frac{q}{mc}A(r,t)\right)\psi(r,t)\right] = -\frac{q}{mc}A(r,t)|\psi(r,t)|^2 \quad (2)$$

where $\psi$ is the wave function of the single super-electron and q is the charge (a single pair wave function has the same property). The supercurrent in the standing wave model is therefore a quantum effect in similarity to the Aharonov-Bohm effect.

The total current is therefore a sum of standing wave currents. The non-locality in the linear relation between J and A comes through the phase difference between these microscopic currents. The non-local kernel is derived in the standing wave model from first principles as a weighted sum over these microscopic currents. In contrast to a standard Bloch function, the standing wave function is a cosine function multiplied by the local atomic wave function. Therefore, we use for the macroscopic current a Pippard-like expression:

$$\mathbf{J}_{total}(\mathbf{r}) = -\frac{e^2}{mc}\int_{B(\mathbf{r},\xi_0)} d\mathbf{r}' \mathbf{A}(\mathbf{r}') K(\mathbf{r}-\mathbf{r}')$$

Where $B(\mathbf{r},\xi_0)$ is a disk around **r** with radius $\xi_0$ and

$$K(\mathbf{r}-\mathbf{r}') = \sum_{\mathbf{k}\in\Omega} \cos^2(\mathbf{k}\cdot(\mathbf{r}-\mathbf{r}')) \quad (3)$$

$\Omega$ is the occupied sub-space of k-space with flat dispersion in the superconducting state.

The standing wave kernel decays therefore as a Dirichlet kernel. Such a kernel, around the point where we measure the current J(**r**), is shown in Fig. 2. The value of $\xi_0$ is easily obtained from the first zero of the Dirichlet kernel. Pippard assumed a linear relation between $1/\xi_0$ and Tc [18]. This linear relation was verified many times in the past. On the basis of this linear relation we can calculate the dependence of Tc on the superfluid density in $La_{2-x}Sr_xCuO_4$ with varying doping.

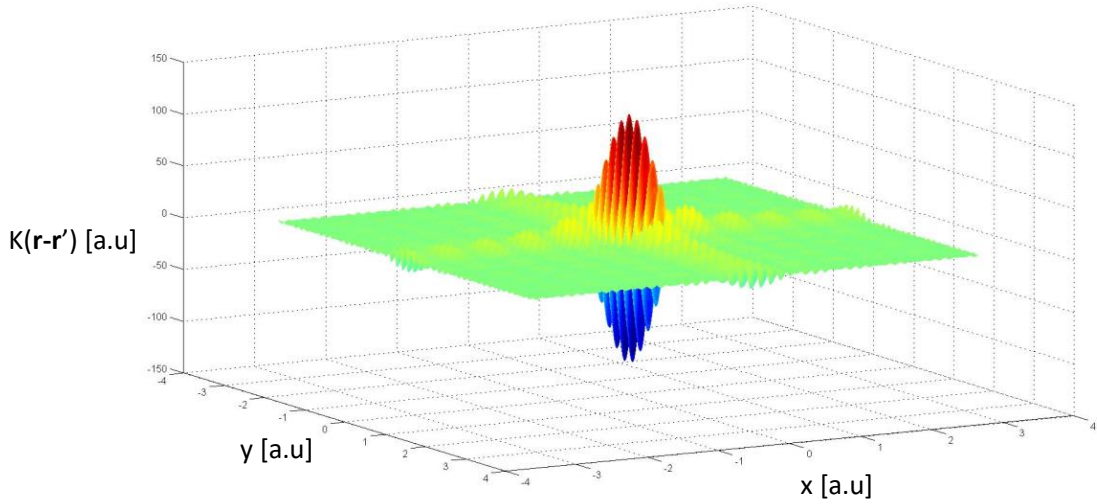

**Fig.2. The standing wave kernel**

The superfluid density is given in the standing wave model as:

$$\rho_0 = \sum_{k \in \Omega} c_k^+ c_k \tag{4}$$

Fig. 3 shows our results. We have calculated the coherence length as function of $\rho_{s0}$ as the shallow regions in k-space are gradually filled with electrons. As the standing wave kernel is not always symmetric in $La_{2-x}Sr_xCuO_4$, we have chosen $\xi_0 = \max(\xi_{0a}, \xi_{0b})$. We then plotted $1/\xi_0$ as function of doping for the over-doped to optimally doped part of the phase diagram in $La_{2-x}Sr_xCuO_4$. The unique behavior in Fig. 3 and in reference 2 can now be explained as originating from the geometry of the Fermi landscape in $La_{2-x}Sr_xCuO_4$. In the highly over-doped region, only a fraction of the shallow region is occupied. This is the low energy part of the blue regions in Fig. 2 right panel. The occupied area (proportional to $\rho_{s0}$ in the standing wave model) in the highly overdoped region can be described as a part of a circle. $1/\xi_0$, derived from the appropriate Dirichlet kernel is proportional to the circle radius, and therefore to the square root of $\rho_{s0}$. As the doping decreases and the number of electrons increases, higher energy regions of the shallow dispersion regions are occupied. These take the form of an elongated shape with one dimension basically fixed. Therefore, the dependence of $1/\xi_0$ on $\rho_{s0}$ (blue dots) is close to linear (red line). Fig. 3 was calculated without any parameter fit. It was only scaled to be compared to the results of reference [1]. The scaling was done by fixing one point at (100,40) to correspond to optimal doping. A better fit can be obtained by more delicate selection of this point. The calculated curve describes the solutions to the standing wave model equations with ARPES data for $La_{2-x}Sr_xCuO_4$ as input. In these calculations, we have used an arbitrary cut-off of 0.03 eV for the square of the absolute value of the gradient of $\varepsilon(\mathbf{k})$ with the wave number $\mathbf{k}$ in the first Brillion zone and the energy $\varepsilon(\mathbf{k})$ at the proximity to the Fermi level. As the experimental results are so close to the calculated function $\xi_0(\rho_{s0})$, even the intersection of the linear fit with the y axis is indistinguishable from Bozovic et al. result of $T_0=7K$.

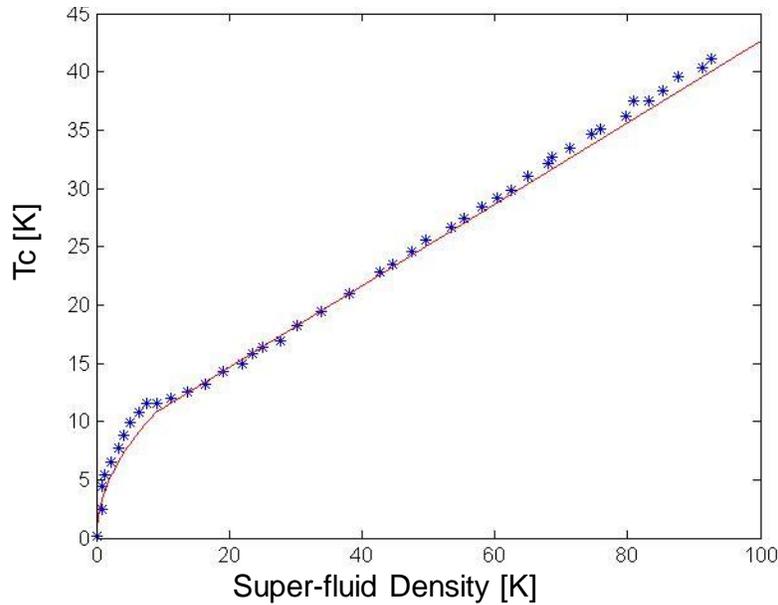

**Fig. 3 The dependence of Tc on the pair density through the New Pippard relation. Blue dots: The measured results of Bozovic et. al. [2]. Red line: calculation based on the standing wave model [1].**

**Summary**

In deriving the calculated curve of Fig. 3 we have used two assumptions, based on the standing wave model:

1. We have used a cutoff on $|\nabla_k \varepsilon|^2$ in accordance with equation (1).
2. We have used the standing wave expression for the single super-electron wave function to derive the kernel K(**r-r′**) and the coherence length $\xi_0$ as its first zero.

**Acknowledgements**

Roger Kornberg is gratefully acknowledged for careful reading and critics of the paper. Alex Demkov is acknowledged for fruitful discussions on the standing wave model. Roald Hoffman is acknowledged for stimulating discussions. Ted Geballe is acknowledged for stimulating discussions.